\documentclass[11pt]{article}
\usepackage[T1]{fontenc}
\usepackage{graphicx}
\usepackage{amsmath,amsfonts,amssymb}
\usepackage{hyperref}
\usepackage{cleveref}
\usepackage{algorithm}
\usepackage{algorithmic}
\usepackage{booktabs}
\usepackage{fancyhdr}
\usepackage{fullpage}

\title{Nash Q-Network for Multi-Agent Cybersecurity Simulation}
\author{Qintong Xie$^{1\star}$, Edward Koh$^{1\star}$, Xavier Cadet$^{1}$, Peter Chin$^{1}$}
\date{} 

\begin{document}
\maketitle

\begin{center}
  $^{1}$Dartmouth College, Hanover, NH 03755\\[1ex]
  $^{\star}$Equal contribution
\end{center}

\begin{abstract}
Cybersecurity defense involves interactions between adversarial parties (namely defenders and hackers), making multi-agent reinforcement learning (MARL) an ideal approach for modeling and learning strategies for these scenarios. This paper addresses one of the key challenges to MARL, the complexity of simultaneous training of agents in nontrivial environments, and presents a novel policy-based Nash Q-learning to directly converge onto a steady equilibrium. We demonstrate the successful implementation of this algorithm in a notable complex cyber defense simulation treated as a two-player zero-sum Markov game setting. We propose the Nash Q-Network, which aims to learn Nash-optimal strategies that translate to robust defenses in cybersecurity settings. Our approach incorporates aspects of proximal policy optimization (PPO), deep Q-network (DQN), and the Nash-Q algorithm, addressing common challenges like non-stationarity and instability in multi-agent learning. The training process employs distributed data collection and carefully designed neural architectures for both agents and critics.

\end{abstract}

\section{Introduction}

Cyber defense and cyber attacks represent inherently adversarial problems characterized by continuous and sophisticated interactions between intelligent parties. These interactions require defenders and attackers to iteratively adapt their strategies in anticipation of their opponent’s behavior. Reinforcement learning (RL) naturally suits this context, as it enables agents to learn optimal decision-making policies by interacting with dynamic environments and adapting to changing opponent behaviors. The success of RL in complex strategy games such as AlphaGo \cite{silver2016mastering} and AlphaStar \cite{vinyals2019grandmaster}, where agents compete against adaptive opponents in high-dimensional, partially observable settings, further highlights its potential to address the challenges of autonomous cyber defense and attack. Consequently, autonomous agents deployed within such environments must exhibit strategic foresight, effectively responding to evolving adversarial tactics.

Conventional RL algorithms typically assume a stationary environment or independent learners optimizing their policies oblivious to the presence and actions of others. While such assumptions may hold in cooperative or single-agent domains, they are often violated in adversarial multi-agent settings such as cybersecurity. Ignoring the strategic reasoning process of opponents can lead to policies that are exploitable or fail to converge, thereby reducing the efficacy of learned defense or attack mechanisms. In light of these limitations, game-theoretic approaches to RL have emerged as a principled means of embedding strategic interaction within the learning process.
Among game-theoretic RL methods, Nash Q-learning \cite{hu2003nash} has garnered attention for its ability to compute policies corresponding to Nash equilibrium solutions of the underlying stochastic game. By explicitly modeling the payoffs and actions of all agents at each state, Nash Q-learning facilitates the identification of equilibrium strategies that reflect mutually best responses. This equilibrium-centric framework enables agents to learn robust policies that are theoretically stable against all adversaries, thereby addressing the strategic complexity inherent in cyber conflict.

In this paper, we introduce the Nash Q-Network (Nash Q-N) in a novel deep reinforcement learning architecture, vastly increasing the representable complexity of the agent and investigate the application of this methodology in a further developed version of the Cyber Operations Research Gym (CybORG) \cite{standen2021cyborg}, a flexible and realistic cybersecurity simulation environment designed to evaluate autonomous strategies for cyber defense and attack. CybORG conceptualizes cyber engagements as multi-agent, stochastic games where the Red Agent embodies an adversary attempting to infiltrate and compromise network assets, while the Blue Agent functions as a defender seeking to detect and mitigate such intrusions. This environment enables the study of strategic interactions under conditions of partial observability and complex system dynamics. 

The principal contributions of this work are as follows:

\begin{itemize}
    \item \textbf{Joint Q-Network Critic:} We formulate a novel joint state and action Q-network enabling us to compute payoff matrices and nash equilibrium policies.
    
    \item \textbf{Nash-Guided Policy Updates:} We train an independent policy network indirectly through its cross entropy divergence from a nash equilibrium calculated from the joint action Q-values at each joint observed state. This gives us a stable policy able to act over each agent's partial observation space.
    
    \item \textbf{Decoupled Optimization:} Our main contribution is the separation of policy and critic updates, which naturally results in a decoupled optimization process. This separation effectively addresses challenges such as partial observability for each agent.
    
\end{itemize}

Through integrating Nash equilibrium computations into our training algorithm, our approach equips autonomous agents to directly converge on stable, equilibrium-based policies and notably defense strategies robust to the range of possible advanced threats. Experimental results demonstrate that the Nash Blue agent consistently improves its policy against the B-line Red adversary, representing a meaningful step forward in developing strategic and adaptive cyber defense.

\section{Related Work}

Nash equilibrium-based methods have long served as foundational tools for modeling strategic interactions in cybersecurity. These methods represent attacker-defender dynamics as games, seeking equilibrium policies where no player can unilaterally improve their payoff, thus ensuring strategic stability and rationality in adversarial settings \cite{littman1994markov,minehart1994markov}. Theoretical guarantees associated with Nash equilibria make them particularly appealing for security analysis and mechanism design. However, classical Nash equilibrium algorithms typically assume static and fully observable environments, which limits their application in realistic cybersecurity contexts characterized by partial observability and stochastic dynamics \cite{roy2010survey}. Furthermore, the exact computation of equilibria becomes infeasible in large-scale or complex games due to high computational complexity \cite{daskalakis2009complexity}.

To overcome these limitations, multi-agent reinforcement learning (MARL) has emerged as a flexible and scalable alternative. MARL enables agents to learn adaptive policies through repeated interactions within dynamic, uncertain environments \cite{busoniu2008comprehensive}. This approach has been successfully applied to various cybersecurity challenges, including intrusion detection \cite{louati2020deep}, adaptive defense strategies \cite{hammad2024deep}, and penetration testing \cite{ghanem2019reinforcement}. Unlike traditional equilibrium methods, MARL naturally accommodates stochasticity and partial observability, though it often encounters challenges related to learning stability, convergence, and lacks explicit equilibrium guarantees.

Integrating Nash equilibrium concepts into MARL constitutes a promising yet relatively underexplored research frontier. Such hybrid approaches seek to combine the theoretical stability of Nash equilibria with the adaptability of data-driven learning agents. Techniques including equilibrium-informed reinforcement learning \cite{brown2020combining} and equilibrium-constrained algorithms\cite{li2024safe} attempt to bridge this gap. However, these approaches are still rare in cybersecurity research, mainly due to challenges such as partial observability, adversarial behavior, and computational complexity.

Initial reinforcement learning applications in cybersecurity predominantly focused on one-sided defense mechanisms, such as intrusion detection through supervised learning or single-agent Q-learning, which often failed to capture the strategic behavior of adaptive attackers \cite{ozkan2021comprehensive}. To tackle this, game-theoretic MARL methods like Nash Q-learning \cite{hu2003nash} have been utilized. Additionally, correlated Q-learning \cite{greenwald2003correlated} employs correlated equilibria to model cooperation or scenarios with partial observability. These methods offer versatile frameworks for modeling attacker-defender interactions including malware propagation, privilege escalation, and lateral movement within networks \cite{alpcan2010network}.

However, as cybersecurity environments grow in complexity and dimensionality—especially with the proliferation of cloud infrastructures and Internet-of-Things (IoT) systems—the scalability limitations of classical Nash Q-learning have become apparent.
Recent advancements have therefore explored the integration of deep learning techniques with Nash Q-learning. Deep Nash Q-learning utilizes neural networks to approximate Q-values and equilibrium strategies within large-scale settings, significantly improving scalability and sample efficiency \cite{casgrain2022deep}. Nevertheless, despite these improvements, challenges remain in balancing computational tractability, partial observability, and dynamic adversarial behavior.

In summary, while Nash Q-learning and its variants provide a principled foundation for strategic learning in cybersecurity, the current work extends this tradition by separating policy and critic updates.

\section{Prerequisite}
\subsection{Stochastic Games}

In a stochastic game, all agents select their actions simultaneously at each time step. Both the state space and action spaces are assumed to be discrete. Formally, an $n$-player stochastic game $\Gamma$ can be defined as a tuple
\[
\langle S, A_1, \ldots, A_n, r_1, \ldots, r_n, p \rangle,
\]
where:
\begin{itemize}
    \item $S$ denotes the set of all possible states,
    \item $A_i$ is the action set available to player $i$, for $i = 1, \ldots, n$,
    \item $r_i : S \times A_1 \times \cdots \times A_n \to \mathbb{R}$ specifies the payoff function for player $i$,
    \item $p : S \times A_1 \times \cdots \times A_n \to \Delta(S)$ defines the state transition probability distribution, where $\Delta(S)$ represents the set of probability distributions over the state space.
\end{itemize}

Given a state \( s \), each agent independently selects an action \( a^1, \ldots, a^n \) and receives a corresponding reward \( r_i(s, a^1, \ldots, a^n) \) for each agent \( i = 1, \ldots, n \). The environment then transitions to a new state \( s' \) according to fixed transition probabilities, which satisfy the normalization constraint
\[
\sum_{s' \in S} p(s' \mid s, a^1, \ldots, a^n) = 1.
\]

In a \emph{discounted stochastic game}, each player’s goal is to maximize the expected sum of discounted rewards, where the discount factor \( \beta \in [0,1) \) controls the importance of future rewards. Denote by \( \pi^i \) the strategy employed by player \( i \). Starting from an initial state \( s \), player \( i \) aims to maximize the value function
\[
v^i(s, \pi^1, \pi^2, \ldots, \pi^n) = \sum_{t=0}^\infty \beta^t \mathbb{E}\left[r^i_t \mid \pi^1, \pi^2, \ldots, \pi^n, s_0 = s\right],
\]
which represents the expected discounted cumulative reward when all players follow their strategies \( \pi^1, \pi^2, \ldots, \pi^n \) starting from state \( s \).

\subsection{Nash Q-Learning}
\label{sec:prerequisites}

Adapting Q-learning to multi-agent settings requires extending the focus from individual actions to joint actions. In a system with $n$ agents, the Q-function for any single agent is expressed as $Q(s, a^1, \ldots, a^n)$, incorporating the actions of all agents, unlike the single-agent case where it depends only on $Q(s,a)$. Building upon this generalized Q-function and utilizing Nash equilibrium as the solution concept, Nash Q-learning \cite{hu2003nash} defined a \emph{Nash Q-value} as the expected cumulative discounted reward for an agent assuming all agents adhere to a joint Nash equilibrium strategy from the subsequent time step onward. This contrasts with single-agent Q-values, where the expectation considers only the agent's own optimal policy.

Agent $i$'s Nash Q-function is defined over the state and joint action space $(s, a^1, \ldots, a^n)$ as the sum of the immediate reward plus the expected discounted future rewards when all agents follow the joint Nash equilibrium strategies. Specifically,
\begin{equation}
Q_*^i(s, a^1, \ldots, a^n) = r^i(s, a^1, \ldots, a^n) + \beta \sum_{s' \in S} p(s' \mid s, a^1, \ldots, a^n) v^i(s', \pi_*^1, \ldots, \pi_*^n),
\end{equation}
where $(\pi_*^1, \ldots, \pi_*^n)$ represent the joint Nash equilibrium strategies; $r^i(s, a^1, \ldots, a^n)$ is the immediate reward received by agent $i$ in state $s$ following joint action $(a^1, \ldots, a^n)$; and $v^i(s', \pi_*^1, \ldots, \pi_*^n)$ denotes agent $i$'s total discounted reward starting from state $s'$ when all agents adhere to these equilibrium strategies.

In scenarios with multiple Nash equilibria, different equilibrium strategy profiles can correspond to distinct Nash Q-functions. 

Nash Q-learning presents an iterative method to compute the equilibria, alternating between two steps: 
\begin{enumerate}
\item Solving the Nash equilibrium for the stage game defined by the current $Q$-functions $\{Q_t\}$ using the Lemke-Howson algorithm \cite{lemke1964equilibrium};
\item Updating the $Q$-value estimates based on the new equilibrium values.
\end{enumerate}

\medskip

This foundational understanding of Nash equilibria and corresponding Q-learning provides critical context and motivation for the algorithmic developments presented in this paper.

\section{Problem Formulation}
\label{sec:problem}
We model the environment as a two-player, zero-sum Markov game between competitive agents \textit{Blue} and \textit{Red}. The game is defined by the tuple $(\mathcal{S}_B, \mathcal{S}_R, \mathcal{A}_B, \mathcal{A}_R, \mathcal{P}, r_B, \gamma)$ where:
\begin{itemize}
    \item $\mathcal{S}_B$ and $\mathcal{S}_R$ are the observation spaces of Blue and Red respectively. At timestep $t$, Blue observes $s_{t,B} \in \mathcal{S}_B$ and Red observes $s_{t,R} \in \mathcal{S}_R$.
    \item $\mathcal{A}_B$ and $\mathcal{A}_R$ are finite action spaces for Blue and Red respectively.
    \item $\mathcal{P}: \mathcal{S}_B \times \mathcal{S}_R \times \mathcal{A}_B \times \mathcal{A}_R \to \Delta(\mathcal{S}_B \times \mathcal{S}_R)$ is the transition function, where $\mathcal{P}(s'_B, s'_R | s_B, s_R, a_B, a_R)$ specifies the probability of transitioning to next states $(s'_B, s'_R)$ given current states $(s_B, s_R)$ and actions $(a_B, a_R)$.
    \item $r_B: \mathcal{S}_B \times \mathcal{S}_R \times \mathcal{A}_B \times \mathcal{A}_R \to \mathbb{R}$ is Blue's reward function. By zero-sum structure, Red's reward is $r_R(s_B, s_R, a_B, a_R) = -r_B(s_B, s_R, a_B, a_R)$.
    \item $\gamma \in [0,1)$ is the discount factor for future rewards.
\end{itemize}
At each timestep $t$:  
\begin{enumerate}
    \item Blue and Red observe their respective state components: $s_{t,B}$ and $s_{t,R}$.  
    \item Agents simultaneously select actions $a_B \in \mathcal{A}_B$ and $a_R \in \mathcal{A}_R$.  
    \item The environment transitions to $(s_{t+1,B}, s_{t+1,R}) \sim \mathcal{P}(\cdot | s_{t,B}, s_{t,R}, a_B, a_R)$.  
    \item Blue receives $r_B(s_{t,B}, s_{t,R}, a_B, a_R)$ and Red receives $-r_B(s_{t,B}, s_{t,R}, a_B, a_R)$.  
\end{enumerate}

Blue's objective is to maximize the expected discounted return:  
\[
\max_{\pi_B} \mathbb{E}\left[ \sum_{t=0}^{\infty} \gamma^t r_B(s_{t,B}, s_{t,R}, a_B^t, a_R^t) \right]
\]  
while Red aims to minimize the same quantity:  
\[
\min_{\pi_R} \mathbb{E}\left[ \sum_{t=0}^{\infty} \gamma^t r_B(s_{t,B}, s_{t,R}, a_B^t, a_R^t) \right]
\]  
where $\pi_B: \mathcal{S}_B \to \Delta(\mathcal{A}_B)$ and $\pi_R: \mathcal{S}_R \to \Delta(\mathcal{A}_R)$ are agents' policies. This defines a fully adversarial Markov game.

\subsection{Nash Equilibrium in Stochastic Games}
Unlike single-agent RL, the multi-agent competitive setup requires solving for Nash equilibria—strategy profiles where no agent can unilaterally improve its expected payoff. Formally, policies $\pi_B^\star, \pi_R^\star$ constitute a Nash equilibrium if:
\[
\begin{cases}
    V_B^{\pi_B^\star, \pi_R^\star}(s_B, s_R) \geq V_B^{\pi_B, \pi_R^\star}(s_B, s_R), \quad \forall \pi_B,\\
    V_R^{\pi_B^\star, \pi_R^\star}(s_B, s_R) \geq V_R^{\pi_B^\star, \pi_R}(s_B, s_R), \quad \forall \pi_R,
\end{cases}
\]
where $V_i^{\pi_B, \pi_R}(s_B, s_R)$ denotes the expected value function for agent $i$ under policies $(\pi_B, \pi_R)$ at state pair $(s_B, s_R)$. This equilibrium concept is the cornerstone of our proposed algorithm.

\section{Methodology}

\subsection{Algorithm Overview}
\label{sec:algorithm-overview}

In adversarial multi-agent reinforcement learning (MARL), agents learn concurrently in non-stationary environments where the behavior of one agent directly influences the learning dynamics of others. This interdependence can lead to instability and convergence challenges if agents optimize policies independently without strategic reasoning about their opponents. 

To overcome these issues, we develop Nash Q-Network (Algorithm \ref{alg:training-framework}) that explicitly incorporates Nash equilibrium computation as a core element of policy optimization within a centralized training framework. The algorithm iterates through three main phases:

\begin{enumerate}
    \item \textbf{Data Collection via Policy Interaction:} Agents simultaneously interact with multiple parallel instances of the environment, sampling actions from their current stochastic policies. This phase generates diverse experience trajectories reflecting evolving strategies.
    
    \item \textbf{Centralized Critic Update Using Policy-Derived Targets:} We employ a centralized critic network that estimates joint state-action value functions. Crucially, instead of using equilibrium strategies for bootstrapping, we compute temporal difference (TD) targets using expectations over the agents' \emph{current} policy distributions. This choice reduces target variance and improves training stability in zero-sum settings where direct equilibrium bootstrapping can induce high variance due to policy oscillations.

    \item \textbf{Decoupled Policy Optimization via Nash Equilibria Alignment:} Policies are updated independently by minimizing divergence from Nash equilibria computed on the joint $Q$-values predicted by the critic. By decoupling policy updates from the critic and anchoring them to fixed equilibrium solutions, we mitigate destabilizing feedback loops and ensure that each agent’s policy evolves toward equilibrium strategies compatible with the latest value estimates.
\end{enumerate}

This synergy between centralized value estimation and equilibrium-aware policy updates enables principled learning of robust strategies in complex multi-agent adversarial domains.

\begin{algorithm}[ht]
\caption{Training Nash Q-Network}
\label{alg:training-framework}
\begin{algorithmic}[1]
\STATE \textbf{Input:} policies $\pi_{\theta_B}, \pi_{\theta_R}$; centralized critic $Q_\phi$
\STATE \textbf{Phase 1:} Collect trajectories by executing $\pi_{\theta_B}, \pi_{\theta_R}$ in parallel environments
\STATE \textbf{Phase 2:} Update critic parameters $\phi$ by minimizing TD error with expected policy-based targets
\STATE \textbf{Phase 3:} Compute Nash equilibria from $Q_\phi(s_B,s_R)$; independently update policies $\pi_{\theta_i}$ to reduce divergence from equilibria
\STATE \textbf{Repeat} until convergence
\end{algorithmic}
\end{algorithm}

\subsection{Agent Architecture}
\label{sec:architecture}

\subsubsection{Policy Network}

Each agent's policy $\pi_{\theta_i}$ is modeled as a parameterized stochastic function approximator producing discrete action distributions conditioned on the observed state $s_i$ ($i \in \{B,R\}$). The policies are realized as multilayer perceptrons (MLPs) capped by a softmax layer, ensuring output vectors form valid probability distributions. Such stochastic policies facilitate efficient exploration by enabling non-deterministic action sampling, which is crucial in strategic adversarial contexts to avoid predictable behavior and exploit opponent weaknesses.

Mathematically, the policy is expressed as:
\[
\pi_{\theta_i}(a_i | s_i) = \operatorname{softmax} \big( \operatorname{MLP}(s_i; \theta_i) \big).
\]

\subsubsection{Rollout Buffer}

To stabilize gradient updates and enhance sample efficiency, each agent maintains a rollout buffer $\mathcal{D}_i$ that stores recent trajectories as tuples
\[
(s_t, a_t, \pi_t, m_t, r_t),
\]
where $m_t$ serves as an action mask, a binary vector encoding which actions are valid in the current state. This mask mechanism is essential in constrained cyber defense environments where some actions may be infeasible or illegal due to system conditions or protocol restrictions. By explicitly applying masks during training and inference, agents learn policies that respect domain constraints and avoid sampling invalid actions.

The replay buffer allows mini-batch sampling to break temporal correlations and promotes robust policy and critic updates with stable gradients.

\subsection{Centralized Critic}
\label{sec:critic}

A hallmark of our approach is the use of a centralized critic network $Q_\phi$ that estimates the joint action-value function over both agents’ actions:
\[
Q_\phi(s_B, s_R, a_B, a_R) \in \mathbb{R}^{|\mathcal{A}_B| \times |\mathcal{A}_R|}.
\]

Unlike decentralized critics that estimate marginal state-action values separately, this centralized structure explicitly models the interaction effects and dependencies between agents’ actions in the adversarial context. The critic receives a joint or concatenated embedding of agent observations and outputs a matrix of Q-values representing the expected returns for all possible joint action pairs at a given pair of states.

This explicit formulation is critical for two reasons:

\begin{itemize}
    \item It enables direct construction of the stage game payoff matrix at each decision step, which is a prerequisite for computing Nash equilibria.
    \item It captures complex strategic dependencies where the value of one agent’s action inherently depends on the counter-actions of its opponent.
\end{itemize}

The Q-network architecture consists of multiple fully connected layers with nonlinear activations, designed to balance expressiveness with computational efficiency.
\subsection{Training Procedure}
\label{sec:training}

\subsubsection{Phase 1: Parallel Data Collection}

To scale data acquisition and accelerate training, we deploy multiple parallel instances of the environment orchestrated via the \texttt{Ray} distributed computing framework \cite{moritz2018ray}. Agents execute their current policies simultaneously, sampling actions from $\pi_{\theta_i}$ while obeying action masks, which reflect environment-specific constraints. This design exploits available hardware concurrency to amass diverse trajectories across many states and strategies.

\subsubsection{Phase 2: Critic Update Using Policy-Based TD Targets}

A critical challenge in multi-agent learning is stabilizing value estimates among evolving opponent policies. We compute TD targets by taking expectations over the agents’ \emph{current} policy distributions rather than directly bootstrapping from equilibrium-based target policies, which can be highly volatile early in training.

Formally, for the next state components $(s'_B, s'_R)$, the bootstrapped target is defined as
\begin{align}
\hat{Q}(s'_B, s'_R) &= \mathbb{E}_{a'_B \sim \pi_{\theta_B}, a'_R \sim \pi_{\theta_R}} \big[ Q_\phi(s'_B, s'_R, a'_B, a'_R) \big] \\
&= \sum_{a_B \in \mathcal{A}_B} \sum_{a_R \in \mathcal{A}_R} \pi_{\theta_B}(a_B\,|\,s'_B) \pi_{\theta_R}(a_R \,|\, s'_R) Q_\phi(s'_B, s'_R, a_B, a_R).
\end{align}

The TD target for time step $t$ is then
\[
y_t = r_t + \gamma \hat{Q}(s_{t+1,B}, s_{t+1,R}),
\]
which provides a smoothed, stable learning signal by averaging over the agents’ policy action distributions. Our critic is trained to minimize the mean squared Bellman error
\[
\mathcal{L}_{\mathrm{critic}} = \mathbb{E}_{(s_{t,B}, s_{t,R}, a_t, r_t, s_{t+1,B}, s_{t+1,R})} \left[ \big( Q_\phi(s_{t,B}, s_{t,R}, a_B, a_R) - y_t \big)^2 \right].
\]

Multiple epochs of gradient descent per batch are used to reduce variance and improve convergence.

\subsubsection{Phase 3: Policy Optimization via Nash Equilibria Alignment}

At each sampled state $(s_B, s_R)$, our method explicitly computes Nash equilibrium mixed strategies $(\sigma_B, \sigma_R)$ over the joint action matrix $Q_\phi(s_B, s_R)$. This Nash solution, representing mutually best-response strategies, provides a principled target distribution for policy improvement.

The policies are updated to minimize the cross-entropy loss aligning policy distributions with computed equilibria:
\[
\mathcal{L}_{\mathrm{policy}}^{i} = -\sum_{a \in \mathcal{A}_i} \sigma_i(a) \log \pi_{\theta_i}(a \mid s_i), \quad i \in \{B,R\},
\]
where for Blue ($i=B$), the policy is conditioned on $s_B$ and for Red ($i=R$), on $s_R$.

In practical scenarios involving constrained action sets, we incorporate action masks into the policy probabilities, ensuring invalid actions receive zero probability and the distribution remains normalized.

This formulation offers several key benefits:
\begin{itemize}
    \item It grounds policy updates in game-theoretically rigorous solutions rather than heuristics or naive best-response approximations.
    \item Decoupling policy updates from the critic’s value estimates stabilizes learning by providing fixed equilibrium targets during each policy update step, diminishing non-stationarity effects.
    \item It encourages policies to evolve toward equilibrium strategies that are robust to adversarial exploitation.
\end{itemize}

Together, these methodological choices enable stable, scalable learning of equilibrium policies in challenging multi-agent cyber defense scenarios.

\section{Implementation Details}
\label{sec:implementation}

\subsection{Neural Network Architectures}

\begin{itemize}
    \item \textbf{Policy Networks:} MLPs with multiple fully connected layers and ReLU activations, terminating in softmax outputs over discrete action sets.
    \item \textbf{Centralized Critic (QMLP):} Outputs a joint $Q$-value matrix over all agent action pairs, explicitly capturing competitive interaction effects. Inputs are concatenated observations from both agents.
\end{itemize}

All networks operate on embedded state vectors consistent with the environment’s observation space.

\subsection{Optimization}

We utilize the Adam optimizer \cite{kingma2014adam} with separate optimizers for policy and critic networks, allowing distinct learning rates tailored to each. Policies minimize cross-entropy divergence with Nash equilibria; the critic minimizes smooth L1 (Huber) loss on TD errors. Hyperparameters such as learning rates and $\beta$ values for Adam are chosen based on empirical tuning to balance convergence speed and stability.

\subsection{Hyperparameters}

\begin{table}[h]
\centering
\caption{Training hyperparameters}
\label{tab:hyperparameters}
\begin{tabular}{ll}
\toprule
Hyperparameter & Value \\
\midrule
Discount factor, $\gamma$ & 0.99 \\
Critic update epochs per batch, $K$ & 6 \\
Rollout horizon, $T$ & 2000 \\
Batch size, $B$ & 64 \\
Policy learning rate & $10^{-3}$ \\
Critic learning rate & $10^{-3}$ \\
Adam betas & (0.9, 0.999) \\
\bottomrule
\end{tabular}
\end{table}

Table \ref{tab:hyperparameters} shows the hyperparameters used in the training process. These values reflect a trade-off between computational feasibility and effective learning in multi-agent adversarial settings.

\section{Experiments and Results}

\subsection{Environment Setting}

We evaluate our proposed Nash Q-learning algorithm within the Cyber Operations Research Gym (CybORG) \cite{standen2021cyborg}, specifically using the Cage Challenge 2 environment \cite{kiely2023autonomous} (CC2). CybORG provides a flexible and realistic simulation platform for autonomous cyber defense research, modeling complex network topologies, vulnerabilities, and adversarial interactions. Prior works have explored various reinforcement learning approaches within CybORG, including hierarchical agent architecture \cite{foley2022autonomous} and differentiable inter agent learning \cite{contractor2024learning}, demonstrating CybORG’s utility as a benchmark for autonomous cyber defense development.

The CC2 environment contains a sophisticated cybersecurity scenario featuring a network under persistent attack by a malicious Red Agent attempting to compromise hosts and spread infection throughout the system. The Blue Agent acts as the autonomous defender, tasked with mitigating intrusions and maintaining network integrity. The environment presents challenges such as partial observability, stochastic system dynamics, and a diverse action space comprising scanning, patching, and isolating hosts.

Our algorithm’s objective in this setting is to minimize the total network infection accrued over episodes, which is reflected through the Blue Agent’s negative reward signal. This reward quantifies the severity and spread of compromises caused by the Red Agent, thereby incentivizing timely and effective defensive actions. The adversarial dynamics in CC2 make it an ideal testbed to assess the ability of multi-agent learning methods to discover robust, equilibrium-based defense strategies.

\subsection{Experiment Workflow}

We first train our NashQ Blue and Red agents for 1000 epochs. Following training, we conduct a comprehensive evaluation where the trained NashQ Blue agents compete against a fixed B-line Red agent—an opponent employing an aggressive exploitation strategy—over 200 evaluation epochs.

Each epoch processes 64 episodes, and each evaluation episode runs for 1000 timesteps with the following metrics recorded per timestep $t$:

\begin{itemize}
    \item \textbf{Cumulative Reward:} The total accumulated reward for the Blue Agent over an episode, representing the overall success in limiting network infection and maintaining system resilience. Higher (less negative) cumulative rewards indicate superior defense. For example, $R_t^B = \sum_{k=0}^t r_k^B$ for Blue agent.
    \item \textbf{Attack Attempts:} $A_t = \sum_{k=0}^t \mathbb{I}[\text{action}_k^R = \text{Impact}]$
    \item \textbf{Successful Impacts:} $I_t = \sum_{k=0}^t \mathbb{I}[\text{action}_k^R = \text{Impact} \land \text{success}_k = 1]$
    \item \textbf{Action Distributions:} Categorical frequencies for both agents:
    \begin{itemize}
        \item Blue: ${\text{Decoys}, \text{Analyse}, \text{Restore}, \text{Block}}$
        \item Red: ${\text{Discovery}, \text{Service Scan}, \text{Exploit}, \text{Privilege Escalate}, \text{Impact}}$
    \end{itemize}
\end{itemize}

Experiments run across 64 parallel episodes using the Ray distributed framework, with results aggregated as:

\begin{align*}
\bar{R}t &= \frac{1}{N} \sum{i=1}^N R_t^{(i)} \\
\sigma_t^R &= \sqrt{\frac{1}{N-1} \sum_{i=1}^N \left(R_t^{(i)} - \bar{R}_t\right)^2}
\end{align*}

\begin{figure}[h]
    \centering
    \includegraphics[width=1\linewidth]{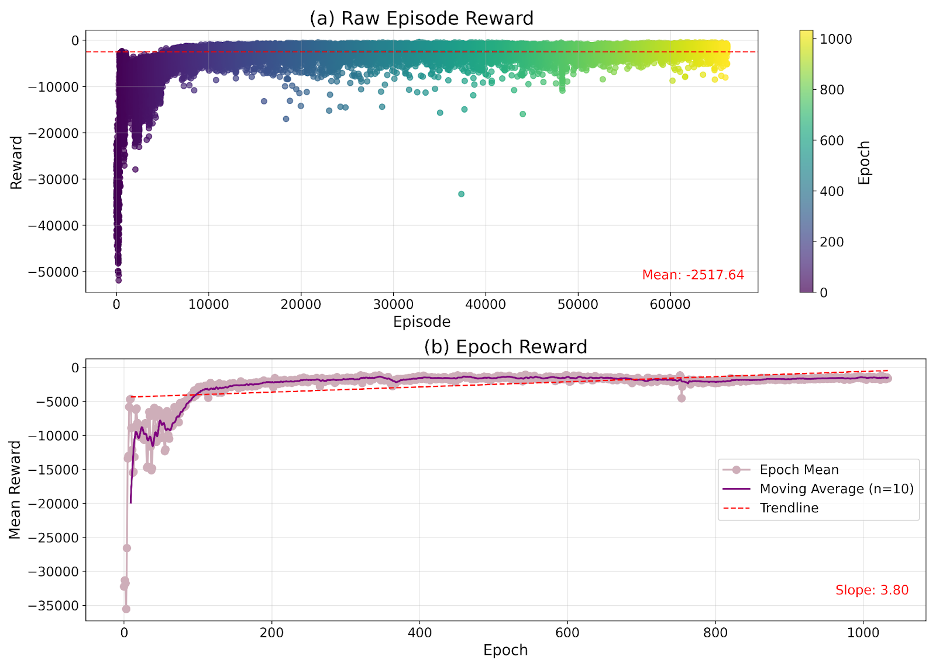}
    \caption{Training performance showing raw episode rewards colored by epoch (a) and epoch reward analysis with moving average and trendline (b). The figure demonstrates a steady improvement in performance over training, with rewards converging and stabilizing after approximately 200 epochs, indicating effective learning and policy refinement.}
    \label{fig:training_performance}
\end{figure}

\begin{figure}
    \centering
    \includegraphics[width=1\linewidth]{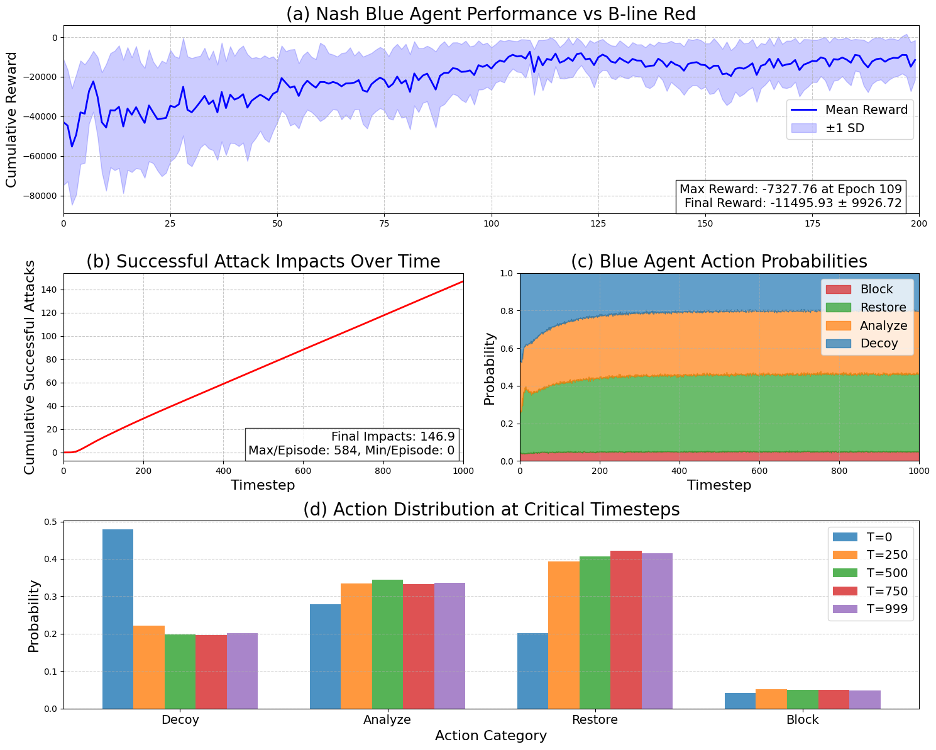}
    \caption{Figure (a) shows steady improvement and convergence in cumulative reward over epochs, indicating effective learning. Figure (b) demonstrates a near-linear increase in successful attack impacts, reflecting the agent’s growing offensive effectiveness within each game. Figure (c) reveals a strategic shift in action probabilities, with \textit{Restore} and \textit{Analyze} actions becoming dominant as training progresses, while \textit{Decoy} actions decline. Figure (d) confirms that at critical timesteps, the agent consistently favors restorative and analytical actions, supporting a stable and focused policy at key moments.}
    \label{fig:performance_overview}
\end{figure}

\subsection{Strategic Effectiveness}
Strategic effectiveness is defined as the mean cumulative reward obtained by the agent at episode termination. This metric evaluates the overall success of the agent's strategy in maximizing long-term returns.

The raw episode rewards (Figure~\ref{fig:training_performance}a) show a large spread initially, including many episodes with substantially negative rewards. However, as training progresses, episode rewards increasingly concentrate near zero, indicating that the agent learns to avoid the worst outcomes and improve cumulative performance. This trend is confirmed by the epoch reward analysis (Figure~\ref{fig:training_performance}b), which shows the per-epoch mean reward increasing over approximately 1000 epochs. The fitted trendline with a positive slope of 3.80 demonstrates consistent performance gain, corroborated by the moving average smoothing that highlights stable improvement beyond initial volatile phases.

Figure~\ref{fig:performance_overview} illustrates the performance of the Nash Blue agent against the B-line Red opponent throughout evaluation. The cumulative reward plot (Figure \ref{fig:performance_overview}a) shows a steady improvement in mean reward over time, with early training epochs exhibiting high variance (indicated by the shaded region representing one standard deviation). Notably, the maximum reward of \(-3665.44\) was obtained at epoch 128, and the final average reward plateaued at approximately \(-5662.69 \pm 4349.63\), suggesting convergence to a relatively stable policy despite some remaining variability.

\subsection{Successful Attack Impacts Over Time}
Successful attack impact measures the cumulative count of the agent’s attacks that have successfully influenced the opponent up to a specific timestep. It reflects the agent's offensive effectiveness and ability to degrade the opponent's state over the course of an episode. 
The cumulative successful attack impacts plot in Figure \ref{fig:performance_overview}b demonstrates a nearly linear increase in the number of successful attacks by the Blue agent over 1000 timesteps. The final count of successful impacts reaches 107.1, with a maximum single-episode impact of 539 and a minimum of zero. This suggests the agent continuously refines offensive capabilities while adapting its defensive actions strategically, as reflected in the evolving action probabilities.

\subsection{Action Selection Probability}
We evaluate the likelihood that the agent selects a particular action at each timestep. These probabilities form a policy distribution over available actions, illustrating the agent's strategic preferences and adaptations as training progresses.

The Blue agent's action selection behavior is explored in Figure \ref{fig:performance_overview}d, which depicts the time evolution of action probabilities and distributions at critical timesteps. Early in training, the agent frequently selects the \emph{Decoy} action, but this probability rapidly decreases as the agent shifts focus toward the \emph{Restore} and \emph{Analyze} actions. By the end of training, \emph{Restore} and \emph{Analyze} dominate the policy, indicating an emphasis on recovery and information gathering strategies.

The stacked area plot in Figure \ref{fig:performance_overview}c confirms this trend quantitatively, showing that \emph{Block} and \emph{Defensive} actions maintain a consistently low selection probability throughout, likely reflecting their limited utility or higher cost. The action distribution at key timesteps (0, 250, 500, 750, 999) further highlights the stabilization of this policy, with the agent nearly abandoning \emph{Decoy} and prioritizing alternatives that presumably yield higher expected returns.

Overall, the experimental results indicate that the Nash Blue agent learns to consistently improve its policy against the B-line Red adversary. Improvements in cumulative rewards and stabilizing action selection patterns reflect effective strategic adaptations. The agent transitions away from initially high-variance behaviors toward more constructive actions such as \emph{Restore} and \emph{Analyze}, enabling sustained attack impacts and defensive resilience. The positive reward trend and convergent behaviors validate the training regime and reward shaping employed in this work.

\section{Conclusion}

In this paper, we propose the Nash Q-Network within adversarial cyber operations to address the inherent strategic complexities of such environments. By modeling cyber engagements as multi-agent stochastic games within the CybORG simulation platform, we demonstrate how the explicit computation of Nash equilibrium strategies informed by joint action Q-values allows agents to anticipate and effectively counter adversarial tactics.

Empirical evaluations indicate that our equilibrium-informed policies consistently outperform heuristic agent baseline, highlighting the importance of embedding strategic reasoning within reinforcement learning frameworks for cybersecurity. By combining the theoretical rigor of Nash equilibria with practical stabilizing techniques, our framework offers a robust foundation for reliable autonomous learning in complex, partially observable, and adversarial multi-agent settings. 

While our study demonstrates promising results, several avenues remain open for further investigation and improvement. One important direction is to extend our framework to scenarios involving more competing agents. Such multi-adversarial environments more closely reflect the complexity of real-world cyber conflicts but also introduce augmented computational challenges. Addressing these requires improving resource efficiency and designing scalable algorithms capable of handling the exponentially larger joint action and observation spaces.

Additionally, we observe that value-based methods relying on Q-functions can be sensitive to sporadic large rewards, leading to instability during training. Notably, architectures inspired by \textit{AlphaStar} incorporate mechanisms to mitigate the adverse effects of such irregular reward signals. Exploring adaptation of these architectural design principles to our cybersecurity domain could enhance learning stability and policy performance.

\section*{Acknowledgments}
This research was funded by the Defense Advanced Research Projects Agency (DARPA), under contract W912CG23C0031.

\bibliographystyle{IEEEtran}
\bibliography{NashQ}

\begin{thebibliography}{10}
\providecommand{\url}[1]{#1}
\csname url@samestyle\endcsname
\providecommand{\newblock}{\relax}
\providecommand{\bibinfo}[2]{#2}
\providecommand{\BIBentrySTDinterwordspacing}{\spaceskip=0pt\relax}
\providecommand{\BIBentryALTinterwordstretchfactor}{4}
\providecommand{\BIBentryALTinterwordspacing}{\spaceskip=\fontdimen2\font plus
\BIBentryALTinterwordstretchfactor\fontdimen3\font minus \fontdimen4\font\relax}
\providecommand{\BIBforeignlanguage}[2]{{%
\expandafter\ifx\csname l@#1\endcsname\relax
\typeout{** WARNING: IEEEtran.bst: No hyphenation pattern has been}%
\typeout{** loaded for the language `#1'. Using the pattern for}%
\typeout{** the default language instead.}%
\else
\language=\csname l@#1\endcsname
\fi
#2}}
\providecommand{\BIBdecl}{\relax}
\BIBdecl

\bibitem{silver2016mastering}
D.~Silver, A.~Huang, C.~J. Maddison, A.~Guez, L.~Sifre, G.~Van Den~Driessche, J.~Schrittwieser, I.~Antonoglou, V.~Panneershelvam, M.~Lanctot \emph{et~al.}, ``Mastering the game of go with deep neural networks and tree search,'' \emph{nature}, vol. 529, no. 7587, pp. 484--489, 2016.

\bibitem{vinyals2019grandmaster}
O.~Vinyals, I.~Babuschkin, W.~M. Czarnecki, M.~Mathieu, A.~Dudzik, J.~Chung, D.~H. Choi, R.~Powell, T.~Ewalds, P.~Georgiev \emph{et~al.}, ``Grandmaster level in starcraft ii using multi-agent reinforcement learning,'' \emph{nature}, vol. 575, no. 7782, pp. 350--354, 2019.

\bibitem{hu2003nash}
J.~Hu and M.~P. Wellman, ``Nash q-learning for general-sum stochastic games,'' \emph{Journal of Machine Learning Research}, vol.~4, pp. 1039--1069, 2003.

\bibitem{standen2021cyborg}
M.~Standen, M.~Lucas, D.~Bowman, T.~J. Richer, J.~Kim, and D.~Marriott, ``Cyborg: A gym for the development of autonomous cyber agents,'' \emph{arXiv preprint arXiv:2108.09118}, 2021.

\bibitem{littman1994markov}
M.~L. Littman, ``Markov games as a framework for multi-agent reinforcement learning,'' in \emph{Machine learning proceedings 1994}.\hskip 1em plus 0.5em minus 0.4em\relax Elsevier, 1994, pp. 157--163.

\bibitem{minehart1994markov}
D.~Minehart and J.~Umanski, ``Markov perfect equilibrium in a repeated principal-agent relationship,'' \emph{Journal of Economic Theory}, vol.~64, no.~2, pp. 322--340, 1994.

\bibitem{roy2010survey}
S.~Roy, C.~Ellis, S.~Shiva, D.~Dasgupta, V.~Shandilya, and Q.~Wu, ``A survey of game theory as applied to network security,'' in \emph{2010 43rd Hawaii international conference on system sciences}.\hskip 1em plus 0.5em minus 0.4em\relax IEEE, 2010, pp. 1--10.

\bibitem{daskalakis2009complexity}
C.~Daskalakis, P.~W. Goldberg, and C.~H. Papadimitriou, ``The complexity of computing a nash equilibrium,'' \emph{Communications of the ACM}, vol.~52, no.~2, pp. 89--97, 2009.

\bibitem{busoniu2008comprehensive}
L.~Busoniu, R.~Babuska, and B.~De~Schutter, ``A comprehensive survey of multiagent reinforcement learning,'' \emph{IEEE Transactions on Systems, Man, and Cybernetics, Part C (Applications and Reviews)}, vol.~38, no.~2, pp. 156--172, 2008.

\bibitem{louati2020deep}
F.~Louati and F.~B. Ktata, ``A deep learning-based multi-agent system for intrusion detection,'' \emph{SN Applied Sciences}, vol.~2, no.~4, p. 675, 2020.

\bibitem{hammad2024deep}
A.~A. Hammad, S.~R. Ahmed, M.~K. Abdul-Hussein, M.~R. Ahmed, D.~A. Majeed, and S.~Algburi, ``Deep reinforcement learning for adaptive cyber defense in network security,'' in \emph{Proceedings of the Cognitive Models and Artificial Intelligence Conference}, 2024, pp. 292--297.

\bibitem{ghanem2019reinforcement}
M.~C. Ghanem and T.~M. Chen, ``Reinforcement learning for efficient network penetration testing,'' \emph{Information}, vol.~11, no.~1, p.~6, 2019.

\bibitem{brown2020combining}
N.~Brown, A.~Bakhtin, A.~Lerer, and Q.~Gong, ``Combining deep reinforcement learning and search for imperfect-information games,'' \emph{Advances in neural information processing systems}, vol.~33, pp. 17\,057--17\,069, 2020.

\bibitem{li2024safe}
Z.~Li and N.~Azizan, ``Safe multi-agent reinforcement learning with convergence to generalized nash equilibrium,'' \emph{arXiv preprint arXiv:2411.15036}, 2024.

\bibitem{ozkan2021comprehensive}
M.~Ozkan-Okay, R.~Samet, {\"O}.~Aslan, and D.~Gupta, ``A comprehensive systematic literature review on intrusion detection systems,'' \emph{IEEE Access}, vol.~9, pp. 157\,727--157\,760, 2021.

\bibitem{greenwald2003correlated}
A.~Greenwald, K.~Hall, R.~Serrano \emph{et~al.}, ``Correlated q-learning,'' in \emph{ICML}, vol.~3, 2003, pp. 242--249.

\bibitem{alpcan2010network}
T.~Alpcan and T.~Ba{\c{s}}ar, \emph{Network security: A decision and game-theoretic approach}.\hskip 1em plus 0.5em minus 0.4em\relax Cambridge University Press, 2010.

\bibitem{casgrain2022deep}
P.~Casgrain, B.~Ning, and S.~Jaimungal, ``Deep q-learning for nash equilibria: Nash-dqn,'' \emph{Applied Mathematical Finance}, vol.~29, no.~1, pp. 62--78, 2022.

\bibitem{lemke1964equilibrium}
C.~E. Lemke and J.~T. Howson, Jr, ``Equilibrium points of bimatrix games,'' \emph{Journal of the Society for industrial and Applied Mathematics}, vol.~12, no.~2, pp. 413--423, 1964.

\bibitem{moritz2018ray}
P.~Moritz, R.~Nishihara, S.~Wang, A.~Tumanov, R.~Liaw, E.~Liang, M.~Elibol, Z.~Yang, W.~Paul, M.~I. Jordan \emph{et~al.}, ``Ray: A distributed framework for emerging $\{$AI$\}$ applications,'' in \emph{13th USENIX symposium on operating systems design and implementation (OSDI 18)}, 2018, pp. 561--577.

\bibitem{kingma2014adam}
D.~P. Kingma and J.~Ba, ``Adam: A method for stochastic optimization,'' \emph{arXiv preprint arXiv:1412.6980}, 2014.

\bibitem{kiely2023autonomous}
M.~Kiely, D.~Bowman, M.~Standen, and C.~Moir, ``On autonomous agents in a cyber defence environment,'' \emph{arXiv preprint arXiv:2309.07388}, 2023.

\bibitem{foley2022autonomous}
M.~Foley, C.~Hicks, K.~Highnam, and V.~Mavroudis, ``Autonomous network defence using reinforcement learning,'' in \emph{Proceedings of the 2022 ACM on Asia Conference on Computer and Communications Security}, 2022, pp. 1252--1254.

\bibitem{contractor2024learning}
F.~Contractor \emph{et~al.}, ``Learning to communicate in multi-agent reinforcement learning for autonomous cyber defence,'' 2024.

\end{thebibliography}

\end{document}